# Multiplexed vector beam conversion via complex structured matter

Runchen Zhang[1,†], Tade Marozsak[1,†], An Aloysius Wang[1,†], Yunqi Zhang[1,†], Yifei Ma[1], Tingxian Gao[2], Haochuan Geng[3], Qihao Han[1], Ben Dai[2], Steve J Elston[1], Stephen M Morris[1, *], and Chao He[1, *]

[1]*Department of Engineering Science, University of Oxford, Parks Road, Oxford, OX1 3PJ, UK*
[2]*Department of Statistics, The Chinese University of Hong Kong, Hong Kong*
[3]*Mathematical Institute, University of Oxford, Woodstock Road, Oxford, OX2 6GG, UK*

[†]These authors contributed equally to this work
*Corresponding author: stephen.morris@eng.ox.ac.uk; chao.he@eng.ox.ac.uk

## Abstract

**Structured light, in which the amplitude, phase, and polarization of an optical field are deliberately tailored in space and time, has enabled unprecedented control over optical fields, paving the way for diverse applications across photonics and optical engineering. However, the prevailing design philosophy, which predominantly focuses on converting a single fixed input into a single desired output, relies on tunability to achieve time-division multiplexing rather than intrinsic design, and is fundamentally incompatible with wavelength-division multiplexing. Here, we propose a general framework for designing structured matter capable of achieving multiple input-output relations simultaneously, thereby enabling passive devices to realize both time-division and wavelength-division multiplexing. Using Stokes skyrmions, which have recently gained attention for their topological properties and promising applications in modern optical communication and computing, as an example, we demonstrate that a simple retarder–diattenuator–retarder cascade can be designed to simultaneously satisfy three arbitrary input-output relations, enabling diverse functionalities within a single passive element. This approach enables complex and multiplexed manipulation of topological numbers, paving the way for high-dimensional on-chip photonic computing based on optical skyrmions.**

## Introduction

Vector beams are a class of structured light characterized by spatially varying polarization across the transverse beam profile and are an important tool in modern photonics that have enabled advanced control over light–matter interactions[1–4], optical information encoding[5–7] and analysis[8–13], high-precision optical trapping[14–16], among many other applications[17,18]. Recently, it has been demonstrated that such vector beams can be engineered to carry topologically protected indices[19–24], making them a promising approach for next-generation optical communication[22,25–28], sensing[29] and computing[30]. It is therefore of great interest to develop strategies that enable the generation, manipulation, and control of vector beams with high flexibility and multiplexing capability.

A particularly promising strategy for the manipulation of vector beams is to design complex, spatially varying structured matter to achieve a given fixed input-output relation. Such structured matter can then be realized using a range of active or passive devices, including q-plates[31], metasurfaces[32,33], gradient-index systems[23,34], glass cones[35], and waveplate arrays[36,37], which can be tailored to meet specific application requirements. However, this core design philosophy is typically restricted to a single input-output relation, which has two important consequences for multiplexing capacity: (1) time-division multiplexing can only be achieved if the medium is tunable and can be modulated in time, and (2) wavelength-division multiplexing is not intrinsically compatible with this design strategy. Thus, to engineer complex, high-functionality devices, it is necessary to adopt a more sophisticated design strategy.

Recent advances in the quality of structured matter point toward a possible starting point for a new strategy, namely, the use of complex cascades with multiple degrees of freedom that can be manipulated to expand functionality and realize multiple input-output relations simultaneously (**Fig 1a**). More specifically, one can take inspiration from machine learning and formulate the $N$ input-output design process as an optimization problem,

$$\min_{p_1,\cdots,p_k} \sum_{i=1}^{N} \iint \left\| f\left(p_1, \cdots, p_k, \lambda_i, u_{in}^i(x,y)\right) - u_{out}^i(x,y) \right\|^2 dxdy$$

which converts the $i$-th input field $u_{in}^i$ (i.e., a field of Stokes or Jones vectors) at wavelength $\lambda_i$ into the desired output field $u_{out}^i$, where $p_1, \cdots, p_k$ denote material parameters of the cascade (i.e., axis orientation and retardance as a function of wavelength for optical retarders, and transmission coefficient for diattenuators, see **Supplementary Note 1**) and $f$ the function describing how the cascade modulates the field.

Notice that the proposed formalism can be applied to any number of input-output relations, although in practice good solutions may not exist if too many relations are specified or if, for example, two very similar inputs are chosen. Our proposed strategy is also inherently compatible with wavelength-division multiplexing, with wavelength directly accounted for in the optimization through the function $f$. In situations involving only linear media, $f$ can be regarded as a composition of Jones or Mueller matrices. However, the formalism is not restricted to this case, as real-world imperfections[38] and nonlinearities[39–41] can also be incorporated through an appropriate choice of $f$. Lastly, in situations where pixel-wise control is employed, the inner integral can be discretized in a straightforward manner.

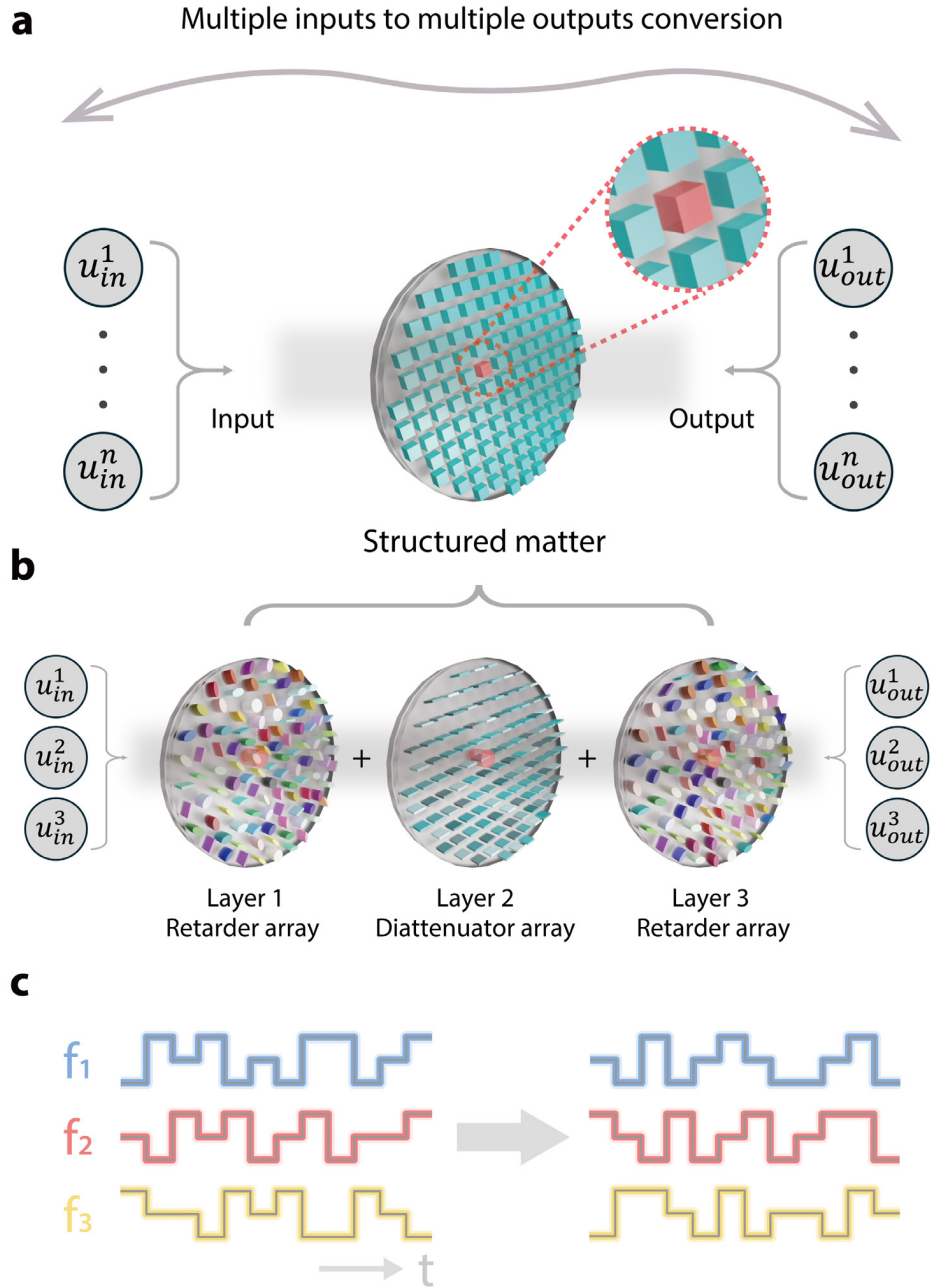

**Fig. 1. Concept. (a)** Complex, spatially varying structured matter can be designed to achieve multiple input-output relations simultaneously, thereby enabling the use of passive components to manipulate structured fields in a manner compatible with both time-division and wavelength-division multiplexing. **(b)** By cascading lower-functionality devices such as optical retarders and diattenuators, complex functionalities can be achieved. As illustrated, a retarder-diattenuator-retarder cascade (see details in main text) can be designed to produce three distinct output polarization fields corresponding to three fixed input polarization fields. In this implementation, the inset pixel in (a) is equivalent to the combined optical modulation of the three highlighted elemental pixels in (b). **(c)** The design supports both time-division multiplexing, where different temporal polarization profiles at a single wavelength are mapped to distinct outputs, and wavelength-division multiplexing, where input polarization profiles at different wavelengths are simultaneously mapped to their respective outputs. As illustrated, the blue, red, and yellow traces represent the temporal signals of $f_1$, $f_2$, and $f_3$, whose time-varying polarization profiles are demonstrated as three discrete levels of each trace. The left side shows the inputs and the right side the transformed outputs. Time-division multiplexing is demonstrated by showing that each individual wavelength ($f_1$, $f_2$, or $f_3$) transforms its time-varying input into a corresponding time-varying output. Wavelength-division multiplexing is demonstrated by showing that independent inputs at $f_1$, $f_2$, and $f_3$ applied at the same moment are simultaneously transformed into their respective outputs.

To demonstrate the effectiveness of our proposed approach, we focus on the design of Stokes skyrmion (see **Supplementary Note 2**) generators and converters using a simple retarder-diattenuator-retarder cascade (**Fig 1b**), and show that a single passive device can produce three distinct arbitrary polarization fields with different skyrmion or generalized skyrmion numbers[42] simply by modulating the input field, in a manner compatible with both time-division and wavelength-division multiplexing (**Fig 1c**). Not only does this represent an important step toward parallelizable, skyrmion-based on-chip photonic computing, but it also marks progress toward more general topological-number-mediated computation and high-density topological information processing.

**Results**

In the context of Stokes skyrmions, their topological properties can be exploited to simplify the proposed design framework when changes in topological number are the only quantity of interest. This is because the topological index of a field remains invariant under continuous deformations, meaning that as long as the loss in the optimization problem is not too large, the functionality of devices designed through the framework will not be affected[30]. More quantitatively, it has been established that the skyrmion numbers of two different fields, $S_1$ and $S_2$, are identical if they satisfy $\|S_1 - S_2\|_\infty < 1$, which allows for a significant margin of error[43]. In this case, the continuity of the field after discretization becomes a more critical factor. Hence, a more suitable optimization for skyrmion related applications is the $H^1$-seminorm regularized problem

$$\min_{p_1,\cdots,p_k} \sum_{i=1}^{N} \max \left\| f\left(p_1, \cdots, p_k, \lambda_i, u_{in}^i(x,y)\right) - u_{out}^i(x,y) \right\| + \kappa \sum_{j=1}^{k} \iint \left\| \nabla p_j(x,y) \right\|^2 dxdy$$

where $\kappa$ is a regularization constant that determines the degree to which sudden discontinuities in the material parameters are suppressed. A more detailed description on how such optimization problems can be solved is given in **Supplementary Note 3**.

The structured matter that we study here is the three-layered cascade shown in **Fig. 1b**: the first layer is a spatially varying retarder with arbitrary fast axes and retardance; the second is a horizontally aligned diattenuator with spatially varying transmission coefficients; and the third is another spatially varying retarder with arbitrary fast axes and retardance. Note that it is typical for the retardance of an optical retarder to be a function of wavelength. Therefore, for both time-division and wavelength-division multiplexing to be achievable within a single device, the designed wavelength range should be sufficiently narrow such that a single effective retardance can accurately represent the device's input–output relations. In this regard, the topological nature of skyrmions is once again advantageous, as the topological number of the output field is expected to remain robust against variations in retardance across a moderate range of wavelengths.

**Fig. 2a-b** shows an example of a skyrmion generator designed using the strategy presented above. In this case, three uniform input fields—0° linear, 45° linear, and right-circular polarization—are converted into conventional Néel-type skyrmions with skyrmion numbers, SkyN = 1, 5, and 10, respectively. Given the simplicity of generating high-quality uniform fields, particularly in on-chip applications[44–47], the proposed design strategy could serve as the foundation for scalable photonic circuits in which the skyrmion number can be precisely engineered and switched through structured-matter design. **Fig. 2c-d** shows an example of a cascade which can convert between fields of different skyrmion numbers. In this example, a standard Néel-type skyrmion of degree 1 is converted into a corresponding Néel-type skyrmion of degree 5, while two distinct skyrmions of degree 5 are transformed into skyrmions of degree 1 and –5, respectively. This demonstrates the complex functionality that can be achieved through a cascade of simple optical components. Note that the single converter shown here represents one example within a broader family of converters that can be designed using the strategy proposed above. Together, this family of converters can be further cascaded to perform complex discrete operations and implement logical transformations, potentially forming key building blocks for next generation topologically protected photonic logic based on optical skyrmions.

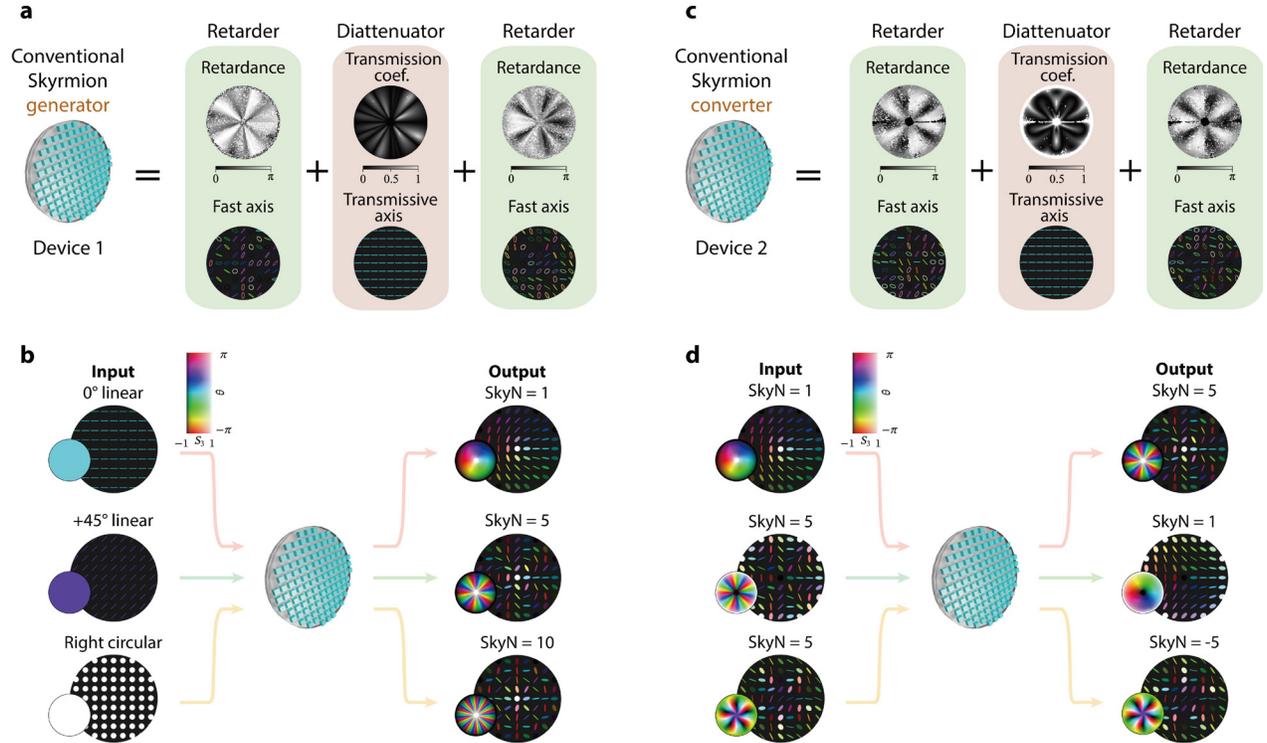

**Fig. 2. Skyrmion generator and converter. (a)** Parameters of the individual layers within the designed cascade used to achieve skyrmion generation, with axis distribution is illustrated by ellipses as in ref. [4]. **(b)** Realized input–output field conversions achieved by the cascade designed in Fig. 2a. Throughout the paper, Stokes fields are visualized using hue to represent the azimuthal angle $\theta$ (defined $\tan\theta = S_2/S_1$) and lightness to represent the $S_3$ component, following ref. [15]. For clarity, the polarization-ellipse distributions of each field are shown. **(c)** Parameters of the individual layers within the designed cascade used to achieve skyrmion conversion. **(d)** Realized input–output field conversions achieved by the cascade designed in Fig. 2c.

The design strategy introduced here is not limited to skyrmion generation and conversion but can also be applied to the manipulation of more complex fields, such as random fields and generalized skyrmions[42] (see **Supplementary Note 2**). **Fig. 3**, for example, demonstrates a device capable of denoising a specific random field into a generalized skyrmion of charge (3,0), as well as performing conversions between different generalized skyrmions. This highlights two important practical implications. The first relates to the conversion of noise into a clean field, which resembles the operation of a generative adversarial network (GAN)[48,49], suggesting that structured-matter design could emulate generative learning processes within the optical domain. The second relates to the capacity for manipulating generalized skyrmions, whose intrinsic data density is significantly higher than that of regular skyrmions[42], and therefore offers another pathway toward high-dimensional optical information encoding and processing.

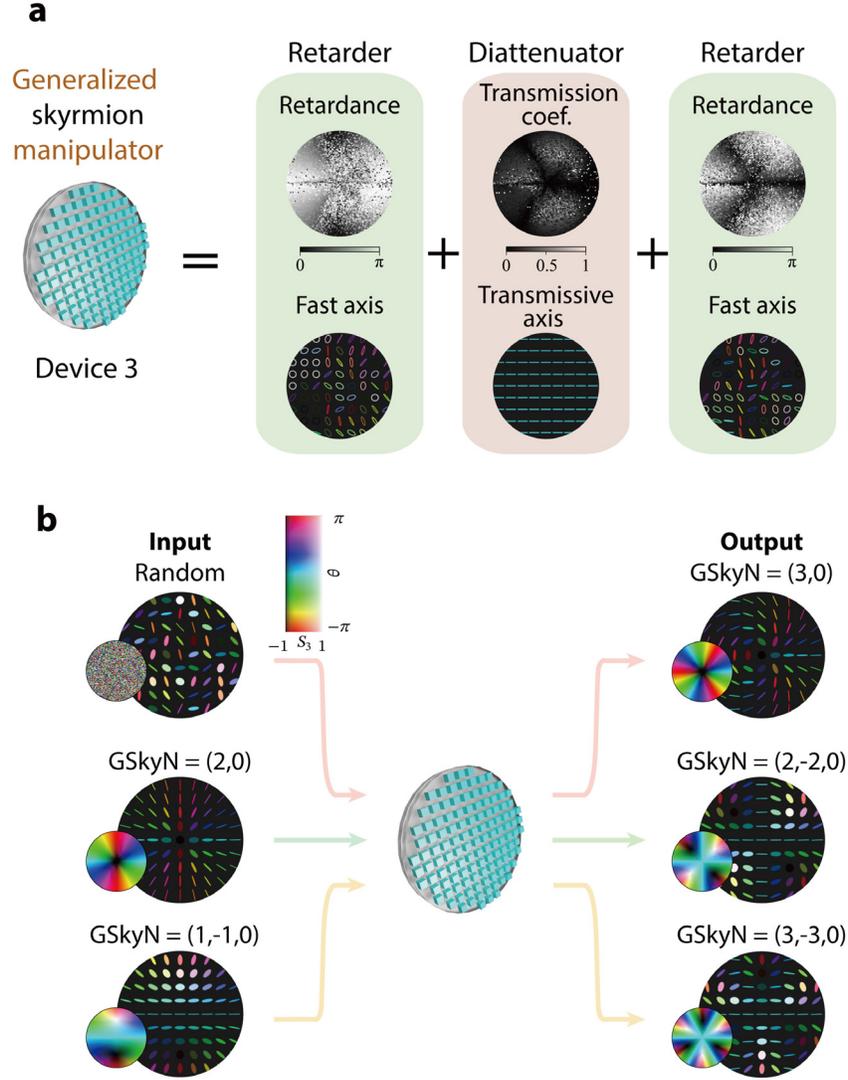

**Fig. 3. Device demonstrating advanced functionalities based on a cascaded structured-matter design.** **(a)** Parameters of the individual layers within the designed cascade used to achieve denoising and generalized skyrmion manipulation. **(b)** Realized input–output field conversions achieved by the cascade designed in Fig. 3a.

Lastly, because the design process introduced relies on an optimization problem, the resulting design does not perfectly produce the desired output field. It is therefore important to quantify the design error to establish that reliable field conversion is achieved. To this end, we define

$$E_{pixel} = \sum_{k}\|S_{tar,k} - S_{out,k}\|^2, \quad S_{out,k} = M_{R_2}M_D\, M_{R_1} S_{in,k}$$

as the pixel error, and its distributions for the three examples in **Figs. 2** and **3** are shown in **Fig. 4**. As illustrated in the figure, although a very small fraction of pixels exhibits errors (see **Supplementary Note 4**) exceeding $10^{-4}$, the vast majority show error levels below $10^{-16}$. These results demonstrate that this family of designs enables highly reliable three-channel multiplexing, which is also the upper limit of the design family (see **Supplementary Note 5**).

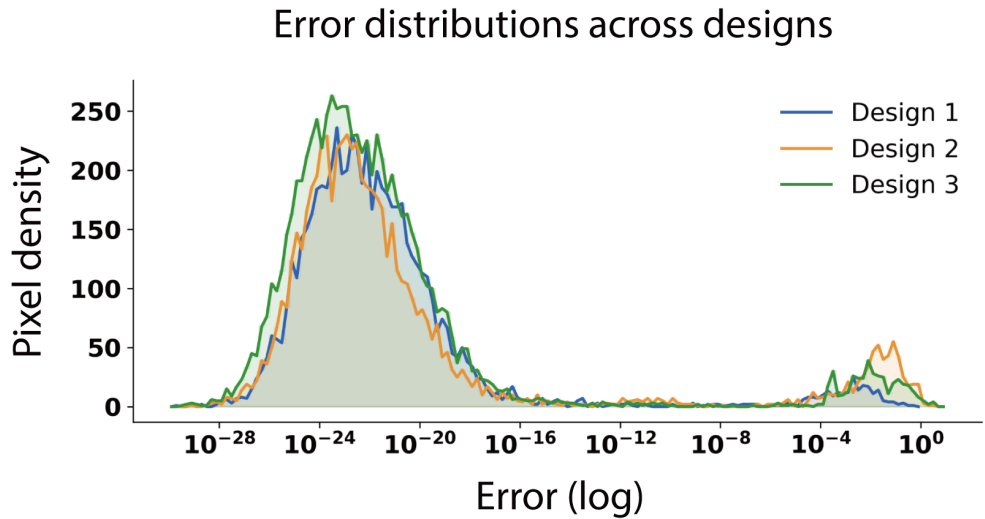

**Fig. 4. Error distributions of the three multiplexing designs.** The distributions of pixel errors for the three examples in Figs. 2 and 3 are shown, with the blue, orange, and green curves corresponding to each example. It can be seen that for all three device designs, although a few pixels exhibit errors exceeding $10^{-4}$, the vast majority have errors within $10^{-16}$, which is approximately machine accuracy, demonstrating that the designs effectively achieve the intended task.

## Discussion

In this work, we proposed a general strategy for designing structured matter that satisfies multiple input–output relations simultaneously, within a framework readily adaptable to a wide range of structured-light applications. When applied to Stokes fields, we demonstrate that a retarder-diattenuator-retarder cascade enables the conversion of three arbitrary input polarization fields into three tailored output polarization fields, thereby realizing diverse functionalities within a single passive element, extending beyond the manipulation of Stokes skyrmions alone.

There are, however, two important points of consideration that can further enhance the applicability of the proposed strategy. The first is a theoretical consideration, namely, the maximum number of input-output relations that can be simultaneously achieved with sufficient precision, and the influence of the cascade structure on the well-posedness of the optimization. A technical analysis of this is provided in **Supplementary Note 4**. The second is a practical consideration, namely, the ability to realize structured matter with the required spatially distributed parameters. In this regard, three additional points are worth noting. First, arbitrary spatially varying retarders[36,37] and diattenuators with spatially varying transmission coefficients[50] have already been experimentally demonstrated, confirming the feasibility of the proposed design. Second, an appropriate choice of cascade structure can simplify fabrication, for example, by employing a diattenuator with a uniformly horizontal transmission axis, as in our proposed design, which is easier to realize than one with spatially varying transmission axes. Third, advances in metasurface technology provide promising routes for realizing smaller and more precise spatially varying retarders[47] and diattenuators[51], which open possibilities for compact, multifunctional optical components capable of implementing complex field transformations.

In conclusion, the proposed design framework enables a single passive device to perform multiple functions that would traditionally require a reconfigurable element or multiple passive components, greatly reducing system complexity and facilitating compact integration. As vector beams continue to demonstrate growing potential in optical communication and computing, we anticipate that this framework will play an important role in advancing next generation integrated photonic technologies.